# Spatio-Temporal Correlation of Epileptic Seizures with The Electrocardiography Brain Perfusion Index

Samuel J. van Bohemen[1], Joe O. Nardo[1], Jeffrey M. Rogers[2], Eleanor Stephens[3], Chong H. Wong[3,4], Andrew F. Bleasel[3,4], Andre Z. Kyme[1,5]

1. The School of Biomedical Engineering, The University of Sydney, Sydney, New South Wales, Australia
2. Department of Clinical Medicine, Macquarie University, Sydney, New South Wales, Australia
3. Department of Neurology, Westmead Hospital, Sydney, New South Wales, Australia
4. Westmead Clinical School, The University of Sydney, Sydney, New South Wales, Australia
5. Brain and Mind Centre, The University of Sydney, Sydney, New South Wales, Australia

**Abstract**

The Electrocardiography Brain Perfusion index (EBPi) is a novel electrocardiography (ECG)-based metric that may function as a proxy for cerebral blood flow (CBF). We investigated the spatio-temporal correlation between EBPi and epileptic seizure events. EBPi was computed retrospectively from clinical EEG and ECG data captured previously from 30 epilepsy patients during seizures. Significant EBPi changes were compared temporally with clinically defined ground-truth seizure onset and offset times. We also assessed the spatial correlation between EBPi metrics and clinically defined ground-truth seizure locations. A significant increase in EBPi was detected 10.5 s [-6, 53] (median [95% confidence interval (CI)]) after ground-truth seizure onset, and a significant decrease in EBPi was detected 5 s [-42, 74] (median [95% CI]) after ground-truth seizure offset. EBPi demonstrated a positive predictive value of 61.5% [33.3, 75] (median [95% CI]) and a sensitivity of 57.1% [38.5, 66.7] (median [95% CI]) for the detection of ground-truth seizure locations. EBPi signals exhibited a temporal sensitivity to seizure events and in some cases were correlated spatially to the seizure location. Therefore, EBPi, which has been linked to CBF, appears to contain spatio-temporal information related to seizure activity and might have useful application in augmenting EEG data captured during seizures.

**Graphical abstract**

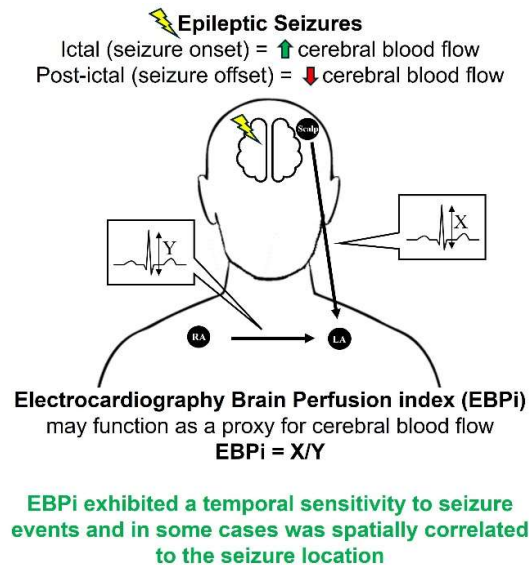

**Graphical Abstract.** Computation of the Electrocardiography Brain Perfusion index (EBPi) during epileptic seizures events.

**Corresponding author:** Samuel J. van Bohemen
**Email:** svan2111@uni.sydney.edu.au
**ORCID ID:** 0000-0001-9620-4099

**Author biographies:**

Mr. Samuel van Bohemen has a first-class honours degree in neuroscience from the University of Otago. He conducted this research as part of his PhD (biomedical engineering) at the University of Sydney.

Mr. Joe Nardo conducted this research as part of his biomedical engineering honours degree at the University of Sydney.

Dr. Jeffrey Rogers has a Bachelor of Science from the University of Michigan, and a Master of Psychology (clinical neuropsychology) and PhD (neuroscience) from the University of Western Australia.

Dr. Eleanor Stephens is a neurologist at Westmead Hospital.

Dr. Chong Wong is a neurologist at Westmead Hospital and The Children's Hospital at Westmead.

Associate Professor Andrew Bleasel is a neurologist at Westmead Hospital and is the Head of the Clinical School at Westmead Hospital.

Dr. Andre Kyme has 20 years of experience developing enabling technologies in medical imaging and clinical applications. He was the senior researcher and author in this work.

**Total word count:** 7519
**Abstract word count:** 200
**Number of figures:** 8
**Number of tables:** 6

## 1. Introduction

Epileptic seizures are associated with clinically significant changes in neuronal activity that can be monitored reliably using electroencephalography (EEG) [1]. Multi-electrode EEG systems tend to be favoured over more involved imaging methods for epilepsy diagnosis, characterisation and monitoring due to their low cost, versatility, non-invasive operation, and because they enable continuous monitoring of neuronal electrical activity [2].

It is well known that the ictal (onset) and post-ictal (offset) stages of epileptic seizures are associated with a respective increase [3] and decrease [4] in cerebral blood flow (CBF) in the affected brain region(s). Numerous studies have reported and characterised seizure-related CBF changes using techniques such as computed tomography perfusion (CTP) imaging [3], magnetic resonance imaging with arterial spin labelling (ASL-MRI) [4], positron emission tomography (PET) [5], single-photon emission computed tomography (SPECT) [6], functional near infrared spectroscopy [7], transcranial Doppler ultrasound [8] and thermal diffusion flowmetry [9]. However, despite the importance of CBF changes surrounding seizure events, CBF measures are not typically used to support seizure detection and epilepsy diagnosis.

Notwithstanding the advantages of EEG over more involved imaging techniques for seizure monitoring, EEG does not provide a direct measure of CBF. Additionally, EEG-based seizure detection and epilepsy diagnosis typically relies on subjective visual inspection of EEG waveforms by an epileptologist. Recent improvements in quantitative EEG using artificial intelligence and machine learning will likely address this shortcoming through sophisticated automated methods that improve reliability and reduce inter-observer variability [10]. Nevertheless, use of EEG in epilepsy will still be limited by the physiological information inherent in the EEG signals.

The Electrocardiography Brain Perfusion index (EBPi) is an electrocardiography (ECG)-based metric which has been linked to changes in CBF [11]. In the recent feasibility study, EBPi was captured using a head-worn device with four EEG scalp electrodes and two ECG chest electrodes using similar hardware to a standard EEG device. Importantly, EBPi can also be computed retrospectively using previously recorded EEG and ECG data provided that at least one scalp electrode and two chest electrodes (LA and RA) were used. This is convenient because it enables existing clinical EEG and ECG data from epilepsy patients to be reprocessed. We exploited this fact in the current work to explore potential clinical benefits of EBPi measurement surrounding seizure events. The key aims of this study were to use multi-electrode EBPi data to: 1) identify optimal change point methods to detect statistically significant fluctuations in EBPi; 2) investigate the temporal correlation between these EBPi changes and ground-truth epileptic seizure onset and offset; and 3) investigate the spatial correlation between EBPi and the ground-truth location of focal seizures.

The overarching aim of this work is not to develop a method to replace EEG for epileptic seizure characterisation and monitoring but rather to explore the potential for EBPi to usefully augment this existing method and provide new or complementary information related to CBF.

## 2. Materials and methods
### 2.1. Raw data

The study used de-identified concurrent EEG and ECG recordings obtained from 30 non-consecutive patients (16 males, 14 females; mean age 35.1 yr, range 20-64 yr, standard deviation (SD) 12.9 yr) presenting with focal epilepsy at the Neurology Department at Westmead Hospital, Sydney, from 2016-19, in accordance with an approved human ethics protocol (WSLHD 2020/ETH01675). All EEG and ECG data were captured on the Neurofax EEG-1200 (Nihon-Kohden Singapore Pte Ltd) recording system at a sampling rate of 500 Hz. EEG data were acquired from 30 scalp electrodes positioned according to the international 10-20 system (Fig. 1), and the concurrent ECG recordings were obtained via 2 chest electrodes at LA and RA (Fig. 2). All data sets were 5-29 min in duration and included a single seizure event based on analysis of the EEG data by an epileptologist. In addition to a presentation of focal epilepsy (see Section 2.2), data sets were selected to have ECG traces with identifiable R-wave peaks and S-wave troughs to enable EBPi computation.

### 2.2. Seizure characterisation

For each seizure event (*n*=30, one per data set), the clinically defined (ground-truth) seizure onset and offset times were determined subjectively by an epileptologist based on visual assessment of changes in the EEG data. Additionally, the ground-truth seizure onset location (as depicted in Fig. 1) was determined based on concordance between seizure semiology, interictal and ictal EEG patterns, magnetic resonance imaging and PET scan data. The concurrent ECG recordings were originally collected for patient monitoring only and had no bearing on this seizure characterisation.

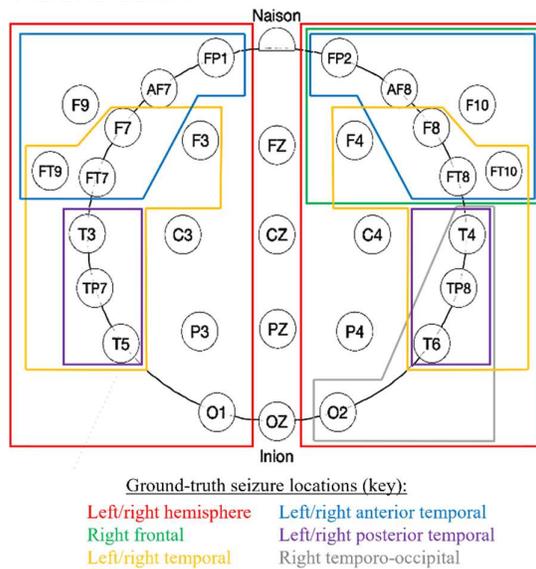

**Fig. 1** Location of the 30 EEG electrodes across the surface of the scalp. Ground-truth seizure locations are overlaid in coloured boxes.

Based on this analysis, all 30 seizures were confirmed as focal seizures according to The International League Against Epilepsy Classification of the Epilepsies [12]. Collectively, the cases in our data set exhibited a pre-seizure interval (before seizure onset) of 306 ± 156 s (mean ± SD), seizure event duration of 102 ± 58 s (mean ± SD) and post-seizure interval (after seizure offset) of 462 ± 198 s (mean ± SD). In 7/30 (23.3%) seizures, the brain region involved in seizure onset could not be identified (termed 'non-localisable'). This was due to EEG artifacts or because the initial EEG change was too subtle or diffuse and could not be localised to specific scalp electrodes.

### 2.3. Electrocardiography Brain Perfusion index (EBPi)

As one of the most electrically conductive components in the body, blood provides a major pathway for the propagation of electrical signals like the ECG [13]. EBPi is based on the assumption that changes in CBF will change the electrical conductivity of blood [14], thus altering the propagation of the ECG signal through the brain. It is measured using the amplitude of the ECG signal recorded across scalp electrodes with respect to the same signal recorded across the chest (Fig. 2). The feasibility of measuring statistically significant changes in EBPi during tasks known to induce changes in CBF was demonstrated in a recent study [11].

EBPi was computed from the seizure data set by re-referencing the EEG data originally recorded at each scalp electrode (Fig. 1) to chest electrode LA (heart adjacent), producing ECG signals at each scalp electrode. We also re-referenced chest electrode RA to chest electrode LA, producing ECG signals across the chest. EBPi was then computed as the ratio of these signals (Fig. 2). Further EBPi details can be found in [11], including examples of the ECG signals shown in Fig. 2.

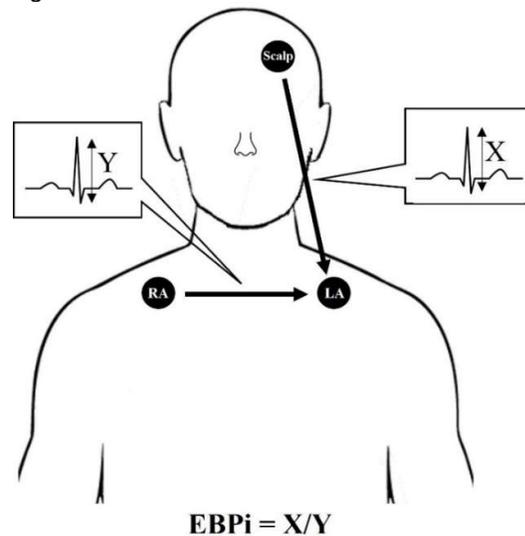

**Fig. 2** Positioning and re-referencing of electrodes to compute EBPi.

EBPi was implemented as in [11] using custom software developed in MATLAB (MathWorks Inc.). For each scalp electrode, EBPi was computed for a given time window $\Delta t$ according to:

$$\text{EBPi}_{k,\Delta t_j} = \frac{\sum_i^{N_j} \frac{A_{k,i}}{A'_i}}{N_j} \qquad (1)$$

where $k$ indexes the scalp electrode, $\Delta t_j$ denotes the *j*-th time interval which contains $N_j$ inlier QRS-complexes indexed by $i$, $A_{k,i}$ is the amplitude difference between the R-wave peak and S-wave trough of the $i$-th QRS-complex for scalp electrode $k$ measured with respect to chest electrode LA, and $A'_i$ is the amplitude of the $i$-th QRS-complex for chest electrode RA measured with respect to chest electrode LA. The time interval $\Delta t$ was set to 20 s to ensure sufficient ECG QRS-complexes were present in each window. This $\Delta t$ corresponds to an EBPi sampling rate of 0.05 Hz. For each seizure event, EBPi values were normalised relative to the mean baseline (pre-seizure) value of the scalp electrode. We define a positive EBPi value as larger than the pre-seizure baseline value, and vice versa for a negative EBPi value.

For each patient, we attempted to compute EBPi for all 30 scalp electrodes for the duration of the recording, which was possible in 24/30 cases. In the remaining cases, occasional artifacts prevented ECG QRS-complexes from being detected in the re-referenced data and resulted in EBPi computation for fewer electrodes (29 electrodes in 5 cases, 28 electrodes in one case).

### 2.4. Detection of statistically significant EBPi change points

The first stage of EBPi analysis was developing a robust method to identify statistically significant changes in the EBPi signal. To detect these statistically significant change points we used a sliding window approach with the studentised permuted Brunner-Munzel (BM) method [15], applied to each scalp electrode. The procedure was as follows:

1. Two contiguous sliding windows of width four samples (80 s, 4x $\Delta t$) were progressed along the time series EBPi data for each scalp electrode and the stochastic equality of the windows was compared using the studentised permuted BM statistical test. Stochastic equality was defined such that the median of window 1 (X) and window 2 (Y) satisfied the probability constraint $P(X<Y)=P(X>Y)$ with a confidence of 95%. Conversely, a change point was defined based on a violation of stochastic equality at the level p<0.05 [15];
2. Identified change points common to at least 10% of the scalp electrodes were retained and the rest were discarded;
3. Retained change points were included in the subsequent temporal correlation analysis (Section 2.5).

### 2.5. Temporal correlation between EBPi change points and ground-truth seizure onset/offset

To investigate the temporal correlation between EBPi change points (Section 2.4) and ground-truth seizure onset/offset, we analysed the temporal distribution of all EBPi change points relative to ground-truth seizure onset and offset times. We also analysed the temporal distribution of the EBPi change points nearest to the ground-truth events (for each seizure). Using the nearest EBPi change points we computed the time difference (s) between these change points and the ground-truth seizure onset and offset times. This analysis was initially performed using EBPi change points regardless of the direction of change (positive or negative). However, we also repeated the analysis considering only positive change points (i.e., significant EBPi increases) for onset, and only negative change points (i.e., significant EBPi decreases) for offset. This was based on the assumption that CBF increases are typically expected at the ictal stage (onset) and decreases at the post-ictal stage (offset).

EBPi change points identified from Section 2.4 were not normally distributed based on the one-sample Kolmogorov-Smirnov test (p<0.05) [16], therefore the temporal correlation analysis was reported using median values and 95% confidence intervals (CI).

### 2.6. Spatial correlation between EBPi response and ground-truth seizure localisation

EEG scalp electrodes associated with the ground-truth seizure locations (depicted in Fig. 1), and the neighbouring EEG scalp electrodes are shown in Table 1.

**Table 1**. Ground-truth seizure locations (depicted in Fig. 1) and the associated and neighbouring (i.e., one EEG scalp electrode position removed) EEG scalp electrodes, for all localisable seizures (*n*=23).

| Ground-truth seizure locations (*n*)† | Associated EEG scalp electrodes | Neighbouring EEG scalp electrodes |
|---|---|---|
| Left hemisphere (2) | Fp1, AF7, F9, F7, F3, FT9, FT7, T3, C3, TP7, T5, P3, O1 | Fp2, Fz, Cz, Pz, Oz |
| Right hemisphere (1) | Fp2, AF8, F10, F8, F4, FT10, FT8, T4, C4, TP8, T6, P4, O2 | Fp1, Fz, Cz, Pz, Oz |
| Right frontal (1) | Fp2, AF8, F10, F8, F4, FT10, FT8 | Fp1, Fz, Cz, C4, T4 |
| Left temporal (6) | F7, F3, FT9, FT7, T3, TP7, T5 | Fp1, AF7, F9, Fz, Cz, C3, P3, O1 |
| Right temporal (2) | F8, F4, FT10, FT8, T4, TP8, T6 | Fp2, AF8, F10, Fz, Cz, C4, P4, O2 |
| Bilateral temporal (1) | F7, F8, F3, F4, FT9, FT10, FT7, FT8, T3, T4, TP7, TP8, T5, T6 | Fp1, Fp2, AF7, AF8, F9, F10, Fz, Cz, C3, C4, P3, P4, O1, O2 |
| Left anterior temporal (3) | Fp1, AF7, F9, F7, FT9, FT7 | Fp2, F3, Fz, C3, T3 |
| Right anterior temporal (1) | Fp2, AF8, F10, F8, FT10, FT8 | Fp1, F4, Fz, C4, T4 |
| Left posterior temporal (4) | T3, TP7, T5 | FT9, FT7, C3, P3, O1 |
| Right posterior temporal (1) | T4, TP8, T6 | FT8, FT10, C4, P4, O2 |
| Right temporo-occipital (1) | T4, TP8, T6, O2 | FT8, FT10, C4, Pz, P4, Oz |

† Number (*n*) of seizures presenting at this brain region.

Spatial localisation of seizure activity was based on the principle of identifying scalp electrodes exhibiting a salient EBPi response during the seizure interval (i.e., between onset and offset). For each scalp electrode, three saliency metrics of EBPi response were used to characterise and compare the EBPi response: 1) area under the EBPi curve (AUC); 2) range of EBPi amplitude; and 3) maximum EBPi amplitude.

For each EBPi saliency metric, *k*-means clustering analysis was used to identify distinct non-overlapping scalp electrode clusters exhibiting the salient response. Gap evaluation was performed to determine the optimal number of scalp electrode clusters (*k*) for each seizure (from *k*=2 to *k*=10, 100 replicates) [17]. Once the optimal number of scalp electrode clusters was determined, *k*-means clustering analysis was performed with 100 replicates to produce robust clusters. The f1-score was used to identify the scalp electrode cluster exhibiting the highest harmonic mean of precision (positive predictive value (PPV)) and sensitivity [18] compared to electrodes associated with the ground-truth seizure location (Table 1).

For each seizure, the scalp electrode cluster identified from the f1-score was compared with three different groups of EEG scalp electrodes to compute PPV and sensitivity: 1) electrodes associated with the ground-truth seizure location; 2) electrodes associated with a slightly dilated region incorporating one extra electrode around the perimeter of the ground-truth region (neighbouring EEG scalp electrodes, see Table 1); and 3) electrodes in the cerebral hemisphere containing the ground-truth region. A true positive (TP) finding represented scalp electrodes implicated in both an EBPi saliency metric and the ground-truth comparison group. A false positive (FP) finding represented scalp electrodes implicated only in an EBPi saliency metric. A false negative (FN) finding represented scalp electrodes implicated only in the ground-truth comparison group. PPV was calculated according to:

$$PPV\ (\%) = \left(\frac{TP}{TP + FP}\right) \times 100 \qquad (2)$$

Sensitivity was calculated according to:

$$Sensitivity\ (\%) = \left(\frac{TP}{TP + FN}\right) \times 100 \qquad (3)$$

The spatial correlation analysis was only implemented for the 23 seizures that had a defined location (i.e., it excluded the 7 seizures considered non-localisable). One seizure was localised to both temporal lobes (bilateral temporal); the hemispheric analysis was not performed for this seizure.

Median PPV and sensitivity were calculated across all seizures. PPV and sensitivity were not normally distributed based on the one-sample Kolmogorov-Smirnov test (p<0.05) [16], therefore the spatial correlation analysis was reported using median values and 95% CI.

## 3. Results
### 3.1. Generation of EBPi data

EBPi data were generated for the 30 seizures in our data set. Three representative EBPi plots before, during and after seizures are displayed in Fig. 3. Plots for all seizures are included in the Supplementary Material Fig. S1.

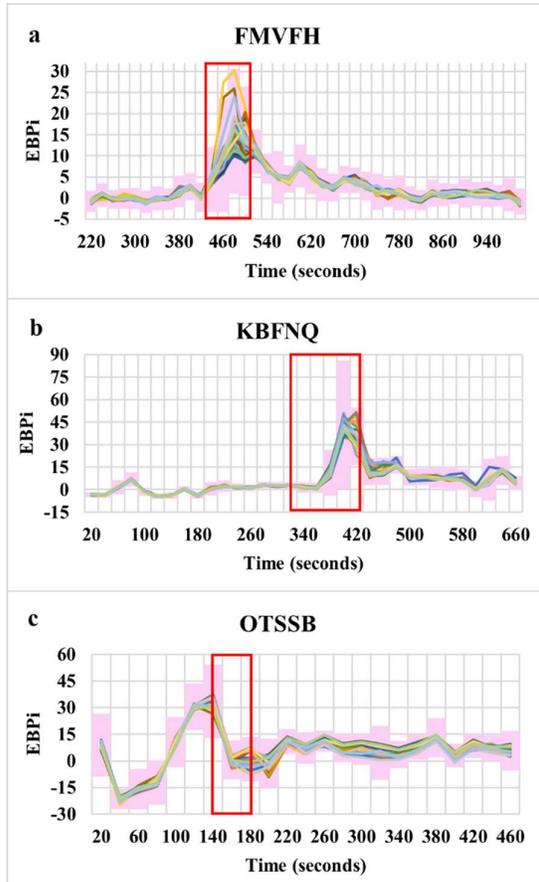

**Fig. 3** Three representative EBPi plots from 3 separate patients before, during and after seizures. Ground-truth seizure onset and offset times are represented by a red box. The different coloured line graphs represent EBPi computed at 30 scalp electrodes. Pink bars represent ± 1 SD. Due to ECG artifacts, EBPi could not be computed during the first 200 s of case FMVFH (3a).

These plots highlight that strong changes in the EBPi signal tended to coincide with seizure events. For example, Fig. 3a shows a strong positive EBPi change coinciding with seizure onset and a negative EBPi change coinciding with seizure offset. Fig. 3b shows a strong positive EBPi change during the seizure and a strong negative EBPi change coinciding with seizure offset. In Fig. 3c several strong EBPi changes are seen to precede and coincide with seizure onset.

### 3.2. Temporal correlation between EBPi change points and ground-truth seizure onset/offset

The temporal distribution of all EBPi change points relative to ground-truth seizure onset, for all seizures, is shown in Fig. 4a. Although change points were detected across a wide time frame, distinct clusters are apparent centred near $t$=0 s, $t$=200 s, $t$=375 s and $t$=550 s, and the density of change points is clearly highest in the approximate range $t$=[-60, 600] s. When only positive change points were considered (Fig. 4b), a similar trend was observed, but the large cluster near t=0 s is more distinct. In Fig. 4c and 4d, rather than histogram all EBPi change points, only the closest change point to onset (blue) and the change point either side (i.e., next earliest (orange) and next latest (yellow)) are considered. The closest change point to onset forms a cluster near t=0 s. This was true regardless of whether we considered change points with either directionality (Fig. 4c) or just positive directionality (Fig. 4d).

**a**     **Temporal distribution of all EBPi change points relative to ground-truth seizure onset**

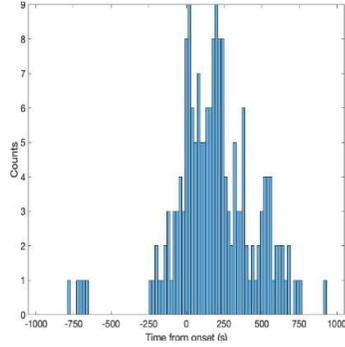

**b**     **Temporal distribution of all positive EBPi change points relative to ground-truth seizure onset**

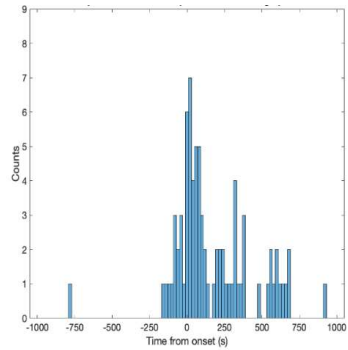

**c**     **Temporal distribution of nearest EBPi change points (either directionality) relative to ground-truth seizure onset**

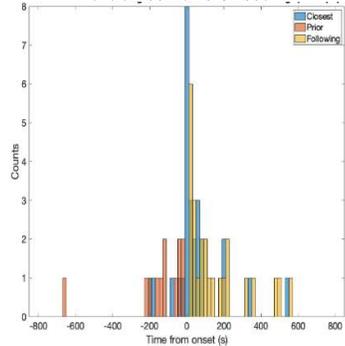

**d**     **Temporal distribution of nearest EBPi change points (only positive) relative to ground-truth seizure onset**

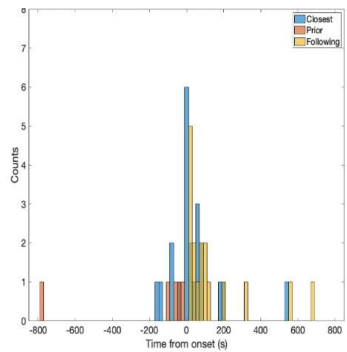

**Fig. 4 (left)** The temporal distribution of all EBPi change points relative to ground-truth seizure onset ($t=0$ s) for all seizures, shown here for change points with either directionality (a) and for only positive change points (b); the temporal distribution of the EBPi change point closest to, prior and following the ground-truth seizure onset for each seizure, shown here for change points with either directionality (c) and only positive change points (d).

The median (signed) time difference between ground-truth seizure onset and the closest EBPi change point is shown in Table 2 for change points of either directionality (Fig. 4c) and for just positive change points (Fig. 4d). The median time difference was approximately 11 s after seizure onset regardless of change point directionality. Both measurements displayed similar confidence intervals.

**Table 2.** Median time difference between ground-truth seizure onset and the closest EBPi change point.

|  | Median time difference (s) | 95% confidence interval (s) |
|---|---|---|
| **Closest EBPi change point** | 12 ($n=27$) [a] | [-5, 50] |
| **Closest positive EBPi change point** | 10.5 ($n=22$) [a] | [-6, 53] |

[a] In some seizures no EBPi change points were detected.

Fig. 5a and 5b show the temporal distribution of all EBPi change points (Fig. 5a) and just the negative change points (Fig. 5b) relative to ground-truth seizure offset, for all seizures. Similar to the onset data (Fig. 4a and 4b), three broad peaks are evident. The broad peak observed at t<0 s represents change points occurring before the ground-truth seizure offset and includes change points near seizure onset (Fig. 4a and 4b). The two broad peaks observed at t>0 s represent change points occurring after the ground-truth seizure offset and is suggestive of a real but systematically shifted EBPi response associated with the post-ictal phase. Again we histogrammed only the closest EBPi change point to offset (blue) and the change point either side (i.e., next earliest (orange) and next latest (yellow)), as shown in Fig. 5c and 5d. The closest change point to offset was typically found in the range $t=[-150, 0]$ s with the largest peak observed near $t=0$ s. This was true when change points of either directionality (Fig. 5c) or just negative change points (Fig. 5b) were considered. These findings suggest there is an EBPi response centred near ground-truth seizure offset.

**Fig. 5 (right)** The temporal distribution of all EBPi change points relative to ground-truth seizure offset ($t$=0 s) for all seizures, shown here for change points with either directionality (a) and for only negative change points (b); the temporal distribution of the EBPi change point closest to, prior and following the ground-truth seizure offset for each seizure, shown here for change points with either directionality (c) and only negative change points (d).

a  **Temporal distribution of all EBPi change points relative to ground-truth seizure offset**

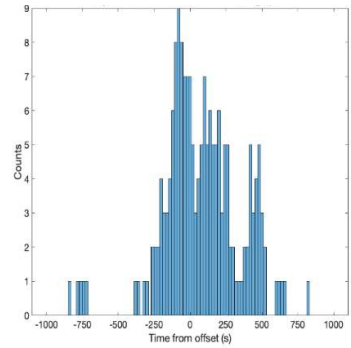

b  **Temporal distribution of all negative EBPi change points relative to ground-truth seizure offset**

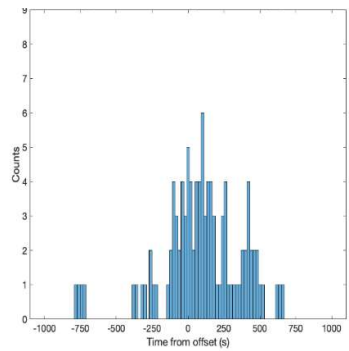

c  **Temporal distribution of nearest EBPi change points (either directionality) relative to ground-truth seizure offset**

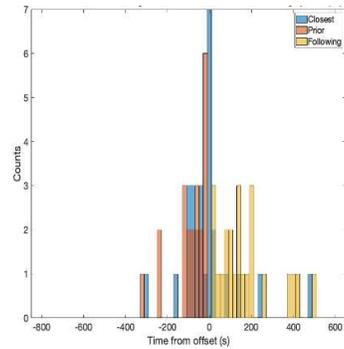

d  **Temporal distribution of nearest EBPi change points (only negative) relative to ground-truth seizure offset**

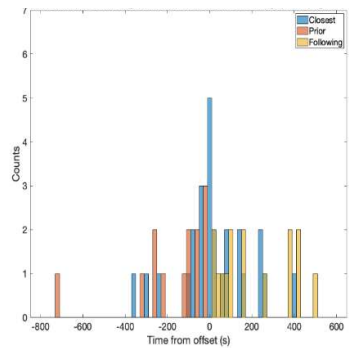

The median (signed) time difference between the ground-truth seizure offset and the closest EBPi change point is shown in Table 3 considering change points of either directionality (Fig. 5c) and just the negative change points (Fig. 5d). When all closest change points were considered (regardless of directionality), the median time difference was 24 s before seizure offset. However, when only the closest negative change points were considered, the median time difference was 5 s after seizure offset. The confidence intervals in both of these measurements were relatively large, indicating that although the EBPi response appears to be temporally correlated with the ground-truth data, there is a 1-2 min range.

**Table 3.** Median time difference between ground-truth seizure offset and the closest EBPi change point.

| | Median time difference (s) | 95% confidence interval (s) |
|---|---|---|
| Closest EBPi change point | -24 ($n$=27) [a] | [-73, 3] |
| Closest negative EBPi change point | 5 ($n$=26) [a] | [-42, 74] |

[a] In some seizures no EBPi change points were detected.

### 3.3. Spatial correlation between EBPi response and ground-truth seizure localisation

The median [95% CI] number of scalp electrode clusters identified for each EBPi saliency metric were as follows: AUC 2 [2, 5], range 3 [2, 4], and amplitude 3 [2, 5]. The median [95% CI] number of scalp electrode clusters across all EBPi saliency metrics was 3 [2, 4].

Fig. 6, 7 and 8 show the spatial correlation between scalp electrodes implicated by the three EBPi saliency metrics and ground-truth seizure locations for the same 3 representative seizures shown in Fig. 3. In these plots, green circles represent scalp electrodes implicated by both an EBPi saliency metric and the ground-truth seizure location (true positive); yellow circles represent scalp electrodes implicated by only the ground-truth seizure location (false negative); red circles represent scalp electrodes implicated by only an EBPi saliency metric (false positive).

For case FMVFH (Fig. 6), EBPi amplitude showed a clear spatial correlation with the ground-truth seizure location. Quantitatively this case showed PPV 100% and sensitivity 75%. EBPi AUC and range demonstrated a moderate spatial correlation with the ground-truth seizure location. EBPi range implicated a distinct electrode (AF7) that was not implicated by the other EBPi saliency metrics, suggesting a possible outlier.

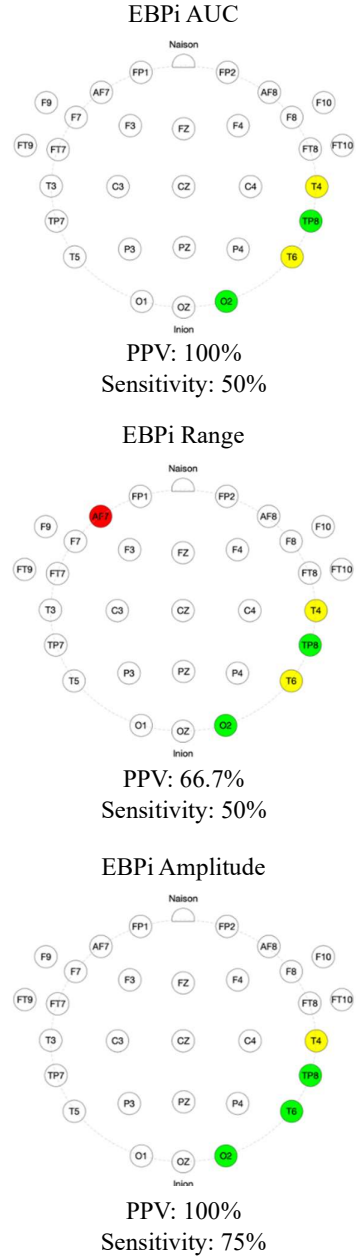

**Fig. 6** Spatial correlation between ground-truth seizure location and EBPi scalp electrodes based on three EBPi saliency metrics, shown here for case FMVFH (cf. Fig. 3a). Green circles represent scalp electrodes implicated by both an EBPi saliency metric and the ground-truth seizure location (true positive); yellow circles represent scalp electrodes implicated by only the ground-truth seizure location (false negative); red circles represent scalp electrodes implicated by only an EBPi saliency metric (false positive). PPV and sensitivity were calculated according to equations (2) and (3).

For case KBFNQ (Fig. 7), EBPi AUC produced a PPV of 100%. EBPi range and amplitude both implicated a group of electrodes (green) common to the ground-truth seizure location, but also more widespread electrodes (red). The sensitivity of all three EBPi metrics was similar (~50-60%).

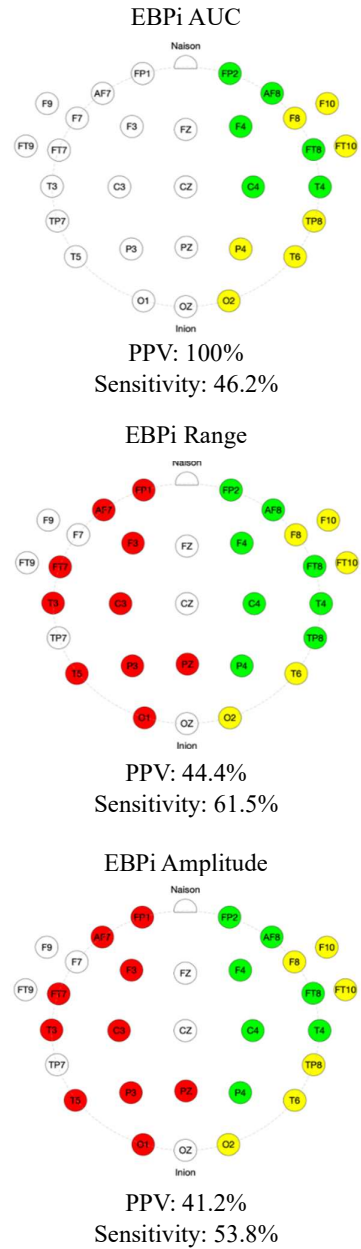

**Seizure KBFNQ EBPi localisation**

EBPi AUC

PPV: 100%
Sensitivity: 46.2%

EBPi Range

PPV: 44.4%
Sensitivity: 61.5%

EBPi Amplitude

PPV: 41.2%
Sensitivity: 53.8%

**Fig. 7** Spatial correlation between ground-truth seizure location and EBPi scalp electrodes based on three EBPi saliency metrics, shown here for case KBFNQ (cf. Fig. 3b). Green circles represent scalp electrodes implicated by both an EBPi saliency metric and the ground-truth seizure location (true positive); yellow circles represent scalp electrodes implicated by only the ground-truth seizure location (false negative); red circles represent scalp electrodes implicated by only an EBPi saliency metric (false positive). PPV and sensitivity were calculated according to equations (2) and (3).

For case OTSSB (Fig. 8), EBPi range produced the highest PPV. EBPi AUC and amplitude implicated a group of electrodes near the ground-truth, with some overlapping (green), and also several electrodes distinct from the ground-truth (red). Both of these EBPi saliency metrics displayed similar PPV and sensitivity.

**Seizure OTSSB EBPi localisation**

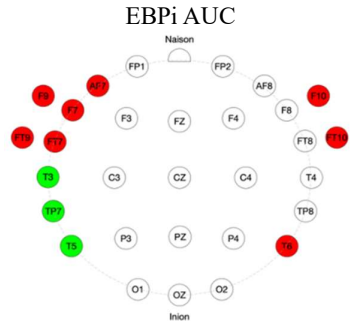

PPV: 27.3%
Sensitivity: 100%

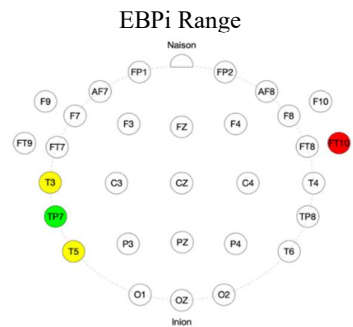

PPV: 50%
Sensitivity: 33.3%

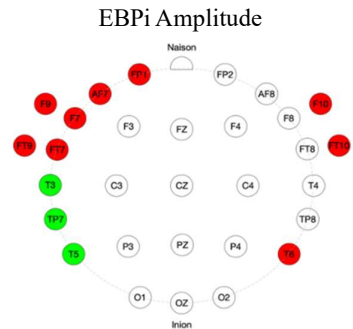

PPV: 25%
Sensitivity: 100%

**Fig. 8** Spatial correlation between ground-truth seizure location and EBPi scalp electrodes based on three EBPi saliency metrics, shown here for case OTSSB (cf. Fig. 3c). Green circles represent scalp electrodes implicated by both an EBPi saliency metric and the ground-truth seizure location (true positive); yellow circles represent scalp electrodes implicated by only the ground-truth seizure location (false negative); red circles represent scalp electrodes implicated by only an EBPi saliency metric (false positive). PPV and sensitivity were calculated according to equations (2) and (3).

Table 4 summaries the median PPV and sensitivity for the three EBPi saliency metrics relative to ground-truth seizure locations. EBPi amplitude demonstrated the highest PPV of 61.5% [33.3, 75] (median [95% CI]) and a sensitivity of 57.1% [38.5, 66.7] (median [95% CI]). All EBPi saliency metrics exhibited similar median and confidence intervals.

Table 4. Median PPV and sensitivity for three EBPi saliency metrics (AUC, range, amplitude) relative to the ground-truth seizure location, for all localisable seizures ($n$=23).

|  | EBPi AUC | EBPi Range | EBPi Amplitude |
|---|---|---|---|
| PPV (median [95% CI]) (%) | 44.4 [28.6, 71.4] | 44.4 [31.3, 60] | 61.5 [33.3, 75] |
| Sensitivity (median [95% CI]) (%) | 57.1 [42.9, 71.4] | 57.1 [38.5, 71.4] | 57.1 [38.5, 66.7] |

Tables 5 and 6 are similar to Table 4, however they relax the comparison to include electrodes inside the ground-truth seizure location and immediately neighbouring this region (Table 5), or electrodes within the same cerebral hemisphere as the ground-truth region (Table 6). This relaxed comparison resulted in higher median PPV (66.7%-83.3%) compared to the strict comparison, as one would expect. There was little difference (<16%) between the two relaxed electrode groupings for both median PPV and sensitivity.

Table 5. Median PPV and sensitivity for three EBPi saliency metrics (AUC, range, amplitude) relative to the ground-truth seizure location and immediately neighbouring EEG scalp electrodes, for all localisable seizures ($n$=23).

|  | EBPi AUC | EBPi Range | EBPi Amplitude |
|---|---|---|---|
| PPV (median [95% CI]) (%) | 71.4 [52.4, 85.7] | 66.7 [50, 80] | 83.3 [52.9, 100] |
| Sensitivity (median [95% CI]) (%) | 40 [26.7, 58.3] | 40 [21.4, 60] | 33.3 [25, 50] |

Table 6. Median PPV and sensitivity for three EBPi saliency metrics (AUC, range, amplitude) relative to the ground-truth seizure hemisphere, for all localisable seizures except one bilateral temporal seizure ($n$=22).

|  | EBPi AUC | EBPi Range | EBPi Amplitude |
|---|---|---|---|
| PPV (median [95% CI]) (%) | 72.1 [54.5, 85.7] | 69 [50, 77.8] | 82.6 [63.6, 100] |
| Sensitivity (median [95% CI]) (%) | 46.2 [30.8, 53.8] | 45.8 [30.8, 61.5] | 38.5 [23.1, 53.8] |

## 4. Discussion

This study was motivated by a recently described ECG-based metric called EBPi which may function as a measurement proxy for CBF [11]. By computing EBPi retrospectively from previously recorded clinical EEG and ECG data, we analysed the temporal and spatial characteristics of EBPi changes with respect to clinically defined focal epileptic seizure events based on multi-modal data including EEG and neuroimaging. Overall, the results suggest that EBPi is sensitive to the ictal and post-ictal stages of epileptic seizures and may also contain information related to the location of focal seizures.

As EBPi is a new measure, there are no standard methods to detect significant EBPi changes. Therefore, an important aspect of this study was the development of an appropriate EBPi change point detection approach. The BM method implemented here, which is suitable for non-parametric data and small sample size (both features of the EBPi data) detected a wide distribution of statistically significant change points, with a predominance of positive and negative change points near ground-truth seizure onset and offset, respectively.

Seizure-induced haemodynamic changes are caused by neurovascular coupling mechanisms to alter CBF in response to changes in electrical and metabolic neuronal activity [19]. Previous studies have reported a close temporal relationship between seizure onset and CBF increases [9] and between seizure offset and CBF decreases [6]. Our results show a wide spread of statistically significant EBPi increases and decreases before, during and after seizure events, but with clear distributions centred around ground-truth seizure onset and offset. Here median values coincided to within approximately 10 s of ground-truth events. Specifically, a significant increase in EBPi was detected 10.5 s [-6, 53] (median [95% CI]) after ground-truth seizure onset, while a significant decrease in EBPi was detected 5 s [-42, 74] (median [95% CI]) after ground-truth seizure offset. Despite typical confidence intervals of 1-2 min, these findings are suggestive that EBPi exhibits a temporal sensitivity to focal seizure events.

Previous studies have reported CTP imaging detected CBF increases in 78.9% of patients during seizure onset [3] and ASL-MRI detected CBF decreases in 92% of patients during seizure offset [20]. Hence most seizures appear to be associated with CBF increases during onset and CBF decreases during offset. Nevertheless, such changes are not always detected and may not always occur. This may explain why significant increases and decreases in EBPi were not detected in all 30 seizures at onset and offset, respectively.

Another important aspect of this study was identifying clusters of scalp electrodes exhibiting a salient EBPi response during seizure intervals and comparing this with ground-truth focal seizure localisation data. In some cases, EBPi saliency metrics exhibited a clear spatial correlation with ground-truth seizure locations (i.e., FMVFH, Fig. 6). However, this was not observed in all cases. Overall, EBPi amplitude demonstrated a strict PPV of 61.5% [33.3, 75] (median [95% CI]) and sensitivity of 57.1% [38.5, 66.7] (median [95% CI]). When the ground-truth regions were relaxed to include neighbouring scalp electrodes, all EBPi saliency metrics demonstrated improved (maximum) PPV but lower sensitivity. The findings were similar when EBPi saliency metrics were compared with the ground-truth hemisphere, suggesting that the EBPi saliency metrics often implicated neighbouring brain regions but were not restricted to a single hemisphere.

Due to the rapid progression of many focal seizures, CBF measurements will often implicate a greater volume of brain than EEG onset localisation. Previous studies have reported CBF dynamics that are partially concordant, fully concordant, or greater than the EEG defined seizure zone [4, 6, 9]. This could explain why EBPi saliency metrics implicated more widespread brain regions compared to the ground-truth seizure locations, which were based on seizure onset only. Furthermore, the accuracy of CBF localisation varies depending on the modality used to measure CBF [3, 20].

Overall, the spatial correlation data we present is suggestive that multi-electrode EBPi data do contain some spatial information related to the seizure. However, further investigation on larger data sets is required to conclude more definitively on the potential accuracy and clinical value of this spatial information.

The timing and location of seizure-induced changes in CBF is likely to be correlated with the magnitude of these changes. EBPi may be less sensitive to smaller changes in CBF, which could explain the modest temporal findings and inconsistent spatial findings. Simultaneous monitoring of CBF using clinically validated methods and EBPi is required to verify the association between EBPi and seizure-induced changes in CBF and to determine what range of CBF changes EBPi is sensitive to.

Although this study indicates that EBPi is responsive to seizure events and may support localising seizures to specific brain regions, there were several important limitations that ought to be considered when interpreting the results.

EBPi is unlikely to be heavily impacted by seizure-induced neuronal electrical activity as EBPi involves the computation of ECG signals at scalp electrodes with amplitudes of ~500-600 μV, whereas seizure activity is ~20-100 μV [21]. Nevertheless, it is not clear how the propagation of the ECG signal through the brain may be affected by seizure-induced electrical activity. Experimental measurements and computational models are required to investigate this further.

We have assumed that both the EEG and ECG electrodes were positioned according to strict clinical guidelines. This is a reasonable assumption with respect to EEG as the placement of these electrodes according to the international 10-20 system is adhered to for seizure monitoring. However, the same cannot be said for the ECG electrodes, as these signals were collected for general patient monitoring and not for seizure detection. This potentially introduced more variability in the EBPi data compared to a strict ECG electrode placement protocol.

Due to the presence of ECG artifacts, in five cases EBPi was only computed for 29 scalp electrodes and in one case for 28 scalp electrodes. In two of these cases the missing scalp electrodes coincided with the ground-truth seizure location, the inclusion of these electrodes may have strengthened the spatial correlation findings; in two cases the missing electrodes did not coincide with the ground-truth seizure location, the inclusion of these electrodes may have weakened the spatial correlation findings; and in one case the seizure was non-localisable, therefore had no impact on the spatial analysis. Also, due to ECG artifacts, EBPi could not be computed for the first 200 s of case FMVFH (Fig. 3a); this was the only case where this occurred. Furthermore, during the seizure events, there were fewer ECG QRS-complexes detected to compute EBPi for each time window ($\Delta t$, equation 1) compared to the pre and post seizure intervals, again due to occasional artifacts. A standardised ECG protocol may have alleviated some of these issues. However, to maximise the reliability of EBPi measurement, robust front-end signal acquisition and/or signal processing methods should be developed to avoid artifacts.

The temporal correlation analysis was reliant on ground-truth seizure onset/offset times determined through visual inspection of EEG data by an epileptologist. Although this is the most commonly used method for seizure detection and epilepsy diagnosis – and the reason why we treated it as the ground truth in this study - this method remains subjective and susceptible to inter-clinician variability [22, 23]. Therefore, this variability is inherent in the strength and confidence of the temporal correlation results we have presented. Alternative methods such as heart rate and heart rate variability can also detect seizure onset and could be utilised in future work [24].

The duration of the seizures varied considerably: mean seizure duration was 102 ± 58 s (mean ± SD, range 41-309 s). Because EBPi was computed for 20 s windows ($\Delta t$, equation 1), there were two seizures during which only two EBPi datapoints were computed at each scalp electrode. A shorter $\Delta t$ would produce more EBPi datapoints, however it would also decrease the number of ECG QRS-complexes detected in each window and increase the likelihood of a noisy or failed EBPi computation. Thus, seizures of longer duration are expected to produce more reliable EBPi data and could be the focus of future studies.

All of the seizures were classified as focal seizures. Further investigation is therefore required to understand EBPi dynamics during other seizure classifications.

In summary, this study provides evidence that EBPi is temporally responsive to focal epileptic seizures and therefore may provide useful complementary information to standard seizure monitoring methods. In turn, it builds on previous work [11] by providing further evidence that EBPi may be sensitive to changes in CBF, since significant EBPi increases and decreases were typically observed at seizure onset and offset, respectively. This study also provides new evidence that EBPi may be sensitive to localised changes in CBF, as in some cases EBPi saliency metrics exhibited a spatial correlation with ground-truth seizure locations. Future EBPi studies using clinically validated CBF methods are required to support these claims. A key advantage of our EBPi approach is that it can be applied retrospectively to conventional EEG data sets and requires no further hardware (apart from concurrent 2-lead ECG). This makes it a very convenient augmentation to existing seizure monitoring based on EEG.

Epileptic seizures are associated with changes in neuronal activity and CBF [3, 4], however CBF measures are not typically used to aid seizure detection or epilepsy diagnosis which is usually based on visual inspection of EEG data. We do not suggest that EBPi could or should replace existing seizure monitoring or detection methods, however because EBPi measures are so easy to obtain alongside existing EEG methods, and since it allows continuous monitoring compared to traditional neuroimaging methods (e.g., PET), it can function as a versatile adjunct to provide potentially complementary information. In future, there is scope to conduct both larger scale retrospective and prospective studies to investigate EBPi measurement alongside EEG measurement during seizures. If clinically validated as a proxy for CBF, EBPi may provide additional pathophysiological information to augment EEG measures of neuronal electrical activity during seizures and other neurological disorders.

## 5. Conclusion

This study investigated the spatio-temporal correlation of EBPi – an ECG-based metric linked to CBF - with focal epileptic seizures. EBPi was temporally responsive to seizure events and in some instances spatially correlated to ground-truth seizure locations. EBPi measurements, which are simple to obtain with existing EEG setups, may be useful to augment EEG-derived seizure information. Additional EBPi studies using clinically validated CBF methods are required to further support EBPi use in this context.

## Supplementary material

See Supplementary Material for EBPi plots before, during and after all 30 seizures (Fig. S1).

## Author contributions

SJVB co-designed the study with AZK, JMR, CHW and AFB. Data were supplied by ES, CHW and AFB. SJVB performed all of the initial data analysis and then refined this analysis with help from JON and AZK. SJVB wrote the manuscript with support from AZK. JON, ES, CHW and AFB edited the manuscript for the final version.

## Conflict of interest

SJVB is a co-founder and shareholder of Nuroflux Pty Ltd. Nuroflux Pty Ltd is the applicant of a published PCT application (WO2021/077154A1) for the described method of monitoring changes in CBF. SJVB and AZK are listed as inventors on the published PCT application.

## Ethics approval

The study was conducted in accordance with an approved human ethics protocol (WSHLD 2020/ETH0167). The study used de-identified EEG and ECG data that had been previously recorded for clinical purposes at Westmead Hospital, Sydney. The study did not involve the recruitment of participants and did not require patient consent.